\documentstyle[aps]{revtex}
\begin{document}
\title{Quantum optical versus quantum Brownian motion master equation
  in terms of covariance and equilibrium properties}
\author{Bassano~Vacchini\footnote{E-mail: bassano.vacchini@mi.infn.it}}
\address{Dipartimento di Fisica
dell'Universit\`a di Milano and INFN, Sezione di Milano,
\\
Via Celoria 16, I-20133 Milan, Italy}
\date{\today}
\maketitle
\begin{abstract}
  Structures of quantum Fokker-Planck equations are characterized with
  respect to the properties of complete positivity, covariance under
  symmetry transformations and satisfaction of equipartition,
  referring to recent mathematical work on structures of unbounded
  generators of covariant quantum dynamical semigroups. In particular
  the quantum optical master equation and the quantum Brownian motion
  master equation are shown to be associated to $\mathrm{U(1)}$ and
  $\mathrm{R}$ symmetry respectively. Considering the motion of a
  Brownian particle, where the expression of the quantum Fokker-Planck
  equation is not completely fixed by the aforementioned requirements,
  a recently introduced microphysical kinetic model is briefly
  recalled, where a quantum generalization of the linear Boltzmann
  equation in the small energy and momentum transfer limit
  straightforwardly leads to quantum Brownian motion.
\end{abstract}
\section{INTRODUCTION}  
\label{I}
A theory of quantum dissipation, even restricted to the Markovian
case, is a subject of major interest for many different scientific
communities, ranging from mathematicians to physicists and chemists,
according to various perspectives. Among these in first line
experimental applications in phenomena where spontaneous emission,
decoherence and dissipation play an important role, but also
theoretical studies regarding the connection between quantum and
classical description of dynamics, since, thanks to the lack of simple
quantization recipes, such as the correspondence principle,
dissipative systems become a fruitful working area where typical
quantum structures may emerge. This interest has led to a huge number
of proposals of Markovian master equations for the description of such
dissipative phenomena, based on microphysical, phenomenological or
purely mathematical approaches (see references
in~\cite{SandulescuIJMPE,Tannor97,Isar99}), not always accompanied by
clear statements with regard to obeyed physical and mathematical
properties, thus often leading to amendments of these models in view
of some missing desired feature. This is in particular true with
respect to the property of positivity or complete
positivity~\cite{Ambegaokar-DiosiEL93-DiosiP-PechukasNATO}, proper
distinction between Hamiltonian and dissipative
part~\cite{Munro-gao-Ford-OC-gao}, translational
invariance~\cite{Ford-OC-traslazioni,reply}, and decoherence
effects~\cite{HuPRA-Kofman}.
\par
As a result some efforts have been made to compare and characterize
the different proposals in view of relevant mathematical and physical
properties: preservation of the positivity of the statistical
operator, existence of a suitable canonical stationary state, and
translational invariance~\cite{Dodonov,Sandulescu,IsarJMP,Tannor97,Yan}. The
main starting point for this research work was the result of Lindblad
for the most general structure of bounded generator of a completely
positive quantum dynamical semigroup~\cite{Lindblad}, together with
his paper on quantum Brownian motion~\cite{LindbladQBM}, in which some
of these issues were already considered.  Complete positivity,
actually equivalent to positivity for the considered Markovian
quasi-free systems~\cite{Talkner}, ensures that the statistical
operator preserves positivity during the time evolution, and has
emerged as a typical quantum feature, corresponding to the requirement
that positivity of the time evolution is preserved under entanglement.
It is in fact by now an essential property in the realm of quantum
communication~\cite{Schumacher-HolevoNEW}, even though its origins lie
in the theory of quantum measurement~\cite{Kraus-Hellwig}.  The result
of Lindblad rigorously holds for bounded operators, though it is
usually exploited as a starting point also for unbounded operators,
leaving in this case the task open, to show that the considered
structure is a proper generator of a completely positive quantum
dynamical semigroup.
\par
In this paper we will try to further clarify the situation, showing
that if besides complete positivity and equipartition, i.e., existence
of a suitable canonical stationary solution, proper covariance
properties of the generator of the dynamics, reflecting the relevant
symmetries of the reservoir (and therefore not necessarily only
translational invariance), are taken into account, a suitable
characterization of two different classes of master equations can be
given in a very neat way.  In particular with the two one-dimensional
Lie groups $\mathrm{U(1)}$ and $\mathrm{R}$ two distinct type of
master equations are associated, describing respectively the damped
harmonic oscillator (the so-called quantum optical master equation)
and the motion of a Brownian particle (the so-called quantum Brownian
motion master equation), which, despite the fact that their physical
realm of validity is essentially well-understood~\cite{GardinerQN},
are sometimes mixed up~\cite{reply}, their characterization in
connection to underlying symmetry of the reservoir being usually
neglected~\cite{GardinerQN}. Recent work on completely positive
quantum dynamical semigroups has shown that, asking for suitable
covariance properties, a characterization of generators of such
semigroups can be given also in the case of unbounded operators, where
very few results are available~\cite{HolevoLNPH}, so that one can
check whether a proposed formal Lindblad structure is indeed a proper
generator of a quantum dynamical semigroup. The room left by these
mathematical and physical requirements should be covered by
microphysical approaches, determining their relevance and predictive
power. It turns out that the quantum optical master equation is in the
essence fixed by these requirements, while in the quantum Brownian
motion case there is a non trivial freedom left, thus explaining the
huge, sometimes contradictory literature devoted to the quantum
Brownian motion master equation. In this connection a recently
obtained~\cite{art3,art4,art6} microphysical model for the quantum
description of the motion of a Brownian particle is presented, derived
from a quantum version of the linear Boltzmann
equation~\cite{art5,art7}, extending previous phenomenological
models~\cite{Joos-Zeh-Gallis90-AlickiPRA} where dissipation effects
leading to the correct stationary solution could not be accounted
for. For further work relying on symmetry properties in the case of a
fermionic oscillator see~\cite{BenattiJPA}.
\par
The paper is organized as follows: in Sec.~\ref{II} we recall the most
general Lindblad structure corresponding to a quantum Fokker-Planck
equation, further showing the expressions that come out asking for
shift-covariance or translation-covariance; in Sec.~\ref{III} we
outline the results of a microphysical model for the description of
quantum Brownian motion, obtained from a quantum linear Boltzmann
equation expressed in terms of the operator-valued dynamic structure
factor of the reservoir; in Sec.~\ref{IV} we briefly comment on our
results, indicating possible future extensions.
\section{GENERAL STRUCTURE OF QUANTUM FOKKER-PLANCK EQUATION WITH A
  LINDBLAD STRUCTURE}
\label{II}
In order to give a quantum description of dissipative phenomena which
at the classical level are described by second-order Fokker-Planck
equations, we first concentrate on generalizations of the quantum
Liouville equation which preserve trace and positivity of the
statistical operator, also possibly accounting for friction effects.
This can be done considering the most general expression of generator
of a completely positive quantum dynamical semigroup (complete
positivity being for these Markovian quasi-free systems necessary in
order to preserve positivity~\cite{Talkner}), given by a Lindblad
structure in which position and momentum operator for the microsystem
(respectively ${\hat {\sf x}}$ and ${\hat {\sf p}}$) appear restricted
to bilinear expressions, according to the classical approximations
leading to a friction term proportional to velocity. The general
expression of a Markovian quantum Fokker-Planck equation preserving
the positivity of the statistical operator is thus given, in the
one-dimensional case to which we will restrict for simplicity,
by~\cite{LindbladQBM,AlbertoQBM,Sandulescu}
\begin{equation}
  \label{l}
  {\cal M}_{{\hat {{\sf x}}}{\hat {{\sf p}}}}[{\hat \rho}]=
        -
        {i\over\hbar}
        \left[
        {H_0 ({\hat {{\sf x}}},{\hat {{\sf p}}})}+\frac{\mu}{2}
        \{{\hat {{\sf x}}},{\hat {{\sf p}}} \} ,{\hat \rho}
        \right]
        +\sum_{i=1}^2
        \left[  
        {\hat {{\sf V}}}_i,
        {\hat \rho}
        {\hat {{\sf V}}}_i^{\dagger} -         \frac 12
        \left \{
        {\hat {{\sf V}}}_i^{\dagger}{\hat {{\sf V}}}_i,
        {\hat \rho}
        \right \}
        \right]  
\end{equation}
\begin{displaymath}
  {\hat {{\sf V}}}_i=\alpha_i{\hat {{\sf p}}} +\beta_i{\hat {{\sf
        x}}}, \qquad \alpha_i,\beta_i\in {\mathrm{C}}, \qquad
  \mu\in\mathrm{R}, 
\end{displaymath}
where $H_0$ is a self-adjoint operator given by a quadratic expression
in ${\hat {{\sf x}}}$ and ${\hat {{\sf p}}}$ describing the free
system, and the added Hamiltonian term proportional to $\mu$ has been
introduced for later convenience.  For the sake of comparison with
classical Fokker-Planck equations, in order to make the intuitive
physical meaning of the different contributions clear, (\ref{l}) is
usually conveniently written in the following form using nested
commutators and anticommutators~\cite{Sandulescu}:
\begin{eqnarray}
  \label{xp}
  {\cal L}_{{\hat {{\sf x}}}{\hat {{\sf p}}}}[{\hat \rho}]
  &=&
        -
        {i\over\hbar}
        [
        {H_0 ({\hat {{\sf x}}},{\hat {{\sf p}}})},{\hat \rho}
        ]
        -
        {i\over\hbar}\frac{(\mu-\gamma)}{2}[\{ {\hat {{\sf x}}},{\hat
          {{\sf p}}} \} ,{\hat \rho}] 
        -
        {i\over\hbar}\gamma [{\hat {{\sf x}}},\{ {\hat {{\sf
              p}}},{\hat \rho} 
        \} ]  
\\
\nonumber
&\hphantom{=}&
        -
        {
        D_{pp}  
        \over
         \hbar^2
        }
        \left[  
        {\hat {{\sf x}}},
        \left[  
        {\hat {{\sf x}}},{\hat \rho}
        \right]  
        \right]  
        -
        {
        D_{xx}
        \over
         \hbar^2
        }
        \left[  
        {\hat {\sf p}},
        \left[  
        {\hat {\sf p}},{\hat \rho}
        \right]  
        \right]  
        +
        {
        D_{px}  
        \over
         \hbar^2
        }
        \left[  
        {\hat {{\sf x}}},
        \left[  
        {\hat {{\sf p}}},{\hat \rho}
        \right]  
        \right]  
        +
        {
        D_{xp}
        \over
         \hbar^2
        }
        \left[  
        {\hat {\sf p}},
        \left[  
        {\hat {\sf x}},{\hat \rho}
        \right]  
        \right]  ,
\end{eqnarray}
where due to $ \left[ {\hat {\sf p}}, \left[ {\hat {\sf x}},\cdot
  \right] \right]= \left[ {\hat {{\sf x}}}, \left[ {\hat {{\sf
          p}}},\cdot \right] \right] $
actually $D_{xp}=D_{px}$ and the new coefficients are related to
$\alpha_i$ and $\beta_i$ through the equations
\begin{displaymath}
  D_{xx}=\frac{\hbar}{2}\sum_{i=1}^2|\alpha_i|^2 \qquad
  D_{pp}=\frac{\hbar}{2}\sum_{i=1}^2|\beta_i|^2 \qquad
  D_{px}=-\frac{\hbar}{2}\Re\sum_{i=1}^2\alpha_i^* \beta_i \qquad
  \gamma=\frac{\hbar}{2}\Im\sum_{i=1}^2\alpha_i^* \beta_i ,
\end{displaymath}
so that the following inequalities hold
\begin{equation}
\label{vincoli}
  D_{xx}\geq 0 \qquad D_{pp}\geq 0 \qquad D_{xx}D_{pp}-D_{px}^2\geq
  \frac{\gamma^2\hbar^2}{4}, 
\end{equation}
which are necessary and sufficient conditions for an expression of the
form (\ref{xp}) to be cast in Lindblad form, corresponding to the
requirement that the matrix of coefficients~\cite{AlbertoQBM}
\begin{equation}
\label{matrix}
 { \mathbf{D}}=\left( \matrix{ D_{xx} & D_{px}+i\frac{\hbar}{2}\gamma
     \cr D_{px}-i\frac{\hbar}{2}\gamma&D_{pp} }\right) 
\end{equation}
has a nonnegative determinant.
\par
An alternative but equivalent expression for (\ref{xp}) can be given
introducing, with the aid of a length $l$, whose physical meaning and
expression will depend on the system to be described, creation and
annihilation operators ${\hat {{\sf a}}}$ and ${{\hat {{\sf
        a}}}^{\scriptscriptstyle\dagger}}$:
\begin{equation}
  \label{relazione}
    {\hat {{\sf a}}}=\frac{1}{l\sqrt{2}} ({\hat {{\sf
          x}}}+\frac{i}{\hbar}l^2 {\hat {{\sf p}}}) 
\qquad
  {{\hat {{\sf a}}}^{\scriptscriptstyle\dagger}}=\frac{1}{l\sqrt{2}}
  ({\hat {{\sf x}}}-\frac{i}{\hbar}l^2 {\hat {{\sf p}}}) ,
\end{equation}
satisfying the commutation relation $[{\hat {{\sf a}}},{\hat
  {{\sf a}}}^{\scriptscriptstyle\dagger}]={\mathrm{1}}$.
One thus obtains the expression:
\begin{eqnarray*}
  {\cal L}_{{\hat {{\sf a}}}{\hat
  {{\sf a}}}^{\scriptscriptstyle\dagger}}[{\hat \rho}]
  &=&
        -
        {i\over\hbar}
        [
        {H_0 ({\hat {{\sf a}}},{\hat
  {{\sf a}}}^{\scriptscriptstyle\dagger})},{\hat \rho}
        ]
        -
        \frac{(\mu-\gamma)}{2}[{\hat {{\sf a}}}^2-{\hat
  {{\sf a}}}^{\scriptscriptstyle\dagger}{}^2 ,{\hat \rho}]    
\\
\nonumber
&\hphantom{=}&
        -\frac{\gamma}{2}
\left[ \vphantom{\frac 12}
[{\hat {{\sf a}}},\{ {\hat {{\sf a}}},{\hat \rho}
        \} ]
-[{\hat {{\sf a}}},\{ {{\hat {{\sf
        a}}}^{\scriptscriptstyle\dagger}},{\hat \rho} 
        \} ]
\right]
+\mathrm{h.c.}
\\
\nonumber
&\hphantom{=}&
        -\frac{1}{2}
\left(\frac{D_{xx}}{l^2}+\frac{D_{pp}l^2}{\hbar^2} \right)
[{{\hat {{\sf a}}}^{\scriptscriptstyle\dagger}},[{\hat {{\sf a}}},{\hat
  \rho}]]
        +\frac{1}{2}
\left(\frac{D_{xx}}{l^2}-\frac{D_{pp}l^2}{\hbar^2}-2i\frac{D_{px}}{\hbar}
\right) 
[{\hat {{\sf a}}},[{\hat {{\sf a}}},{\hat \rho}]]
+\mathrm{h.c.}
,
\end{eqnarray*}
or equivalently, collecting terms as in (\ref{l})
\begin{eqnarray}
\label{aa}
  {\cal M}_{{\hat {{\sf a}}}{\hat
  {{\sf a}}}^{\scriptscriptstyle\dagger}}[{\hat \rho}]
  &=&
        -
        {i\over\hbar}
        [
        {H_0 ({\hat {{\sf a}}},{\hat
  {{\sf a}}}^{\scriptscriptstyle\dagger})},{\hat \rho}
        ]
        -
        \frac{\mu}{2}[{\hat {{\sf a}}}^2-{\hat
  {{\sf a}}}^{\scriptscriptstyle\dagger}{}^2 ,{\hat \rho}]  
\\
\nonumber
&\hphantom{=}&
        +
\left(\frac{D_{xx}}{l^2}+\frac{D_{pp}l^2}{\hbar^2}+\gamma \right)
\left[ 
{\hat {{\sf a}}}{\hat
  \rho}{{\hat {{\sf a}}}^{\scriptscriptstyle\dagger}} -         \frac 12
        \left \{
                {\hat \rho},{{\hat {{\sf
                        a}}}^{\scriptscriptstyle\dagger}}{\hat {{\sf
                      a}}} 
        \right \}
\right]
\\
\nonumber
&\hphantom{=}&
        +
\left(\frac{D_{xx}}{l^2}+\frac{D_{pp}l^2}{\hbar^2}-\gamma \right)
\left[ 
{{\hat {{\sf a}}}^{\scriptscriptstyle\dagger}}{\hat
  \rho}{\hat {{\sf a}}} -         \frac 12
        \left \{
                {\hat \rho},{\hat {{\sf a}}}{{\hat {{\sf
                        a}}}^{\scriptscriptstyle\dagger}} 
        \right \}
\right]
\\
\nonumber
&\hphantom{=}&
        -
\left(\frac{D_{xx}}{l^2}-
\frac{D_{pp}l^2}{\hbar^2}-2i\frac{D_{px}}{\hbar}
\right) 
\left[ 
{\hat {{\sf a}}}{\hat
  \rho}{\hat {{\sf a}}} -         \frac 12
        \left \{
                {\hat \rho},{\hat {{\sf a}}}^2
        \right \}
\right]
+\mathrm{h.c.}
\end{eqnarray}
The recalled expressions essentially give the possible Lindblad
structures at most bilinear in the operators ${\hat {\sf x}}$ and
${\hat {\sf p}}$ or ${\hat {{\sf a}}}$ and ${\hat {{\sf
      a}}}^{\scriptscriptstyle\dagger}$.
\subsection{Shift-covariance}
\label{IA}
We now analyze the behavior of the considered expressions with respect
to suitable symmetry transformations. Consider a locally-compact group
$G$ and a unitary representation $\hat{U} (g)$, with $g \in G$, on the
Hilbert space of the system: following~\cite{HolevoJMP} we say that a
mapping ${\cal F}$ in the Schr\"odinger picture is $G$-covariant if it
commutes with the mapping ${\cal U}_g [\cdot]=\hat{U} (g) \cdot
\hat{U}^{\dagger} (g)$ for all $g \in G$
\begin{equation}
  \label{100}
  {\cal F}[{\cal U}_g[\cdot]]={\cal U}_g[{\cal F}[\cdot]]. 
\end{equation}
\par
Let us now consider the following unitary representation of the
group $\mathrm{U(1)}$
\begin{displaymath}
  \hat{U}_\phi=e^{i\phi \hat{\sf N}}
\end{displaymath}
where $\hat{\sf N}={{\hat {{\sf a}}}^{\scriptscriptstyle\dagger}}{\hat
  {{\sf a}}}$ is the number operator and $\phi\in[0,2\pi]$. If we now
ask for the general expression (\ref{aa}) invariance under the action
of the group $\mathrm{U(1)}$, i.e., shift-covariance according
to~\cite{HolevoJFA-HolevoRMP33}
\begin{equation}
  \label{1}
  {\cal M}_{{\hat {{\sf a}}}{\hat
  {{\sf a}}}^{\scriptscriptstyle\dagger}}[e^{i\phi \hat{\sf N}}\cdot
e^{-i\phi \hat{\sf N}}]=e^{i\phi \hat{\sf N}}{\cal M}_{{\hat 
      {{\sf a}}}{\hat 
  {{\sf a}}}^{\scriptscriptstyle\dagger}}[\cdot]e^{-i\phi \hat{\sf N}} ,
\end{equation}
then the Hamiltonian has to be a function of the generator of the
transformation $\hat{\sf N}$, and the following stringent requirements
appear:
\begin{displaymath}
  D_{xx}=D_{pp}\frac{l^4}{\hbar^2}\qquad D_{px}=0\qquad \mu=0.
\end{displaymath}
Note that the condition $\mu=0$ appears here as a necessary condition
for shift-covariance or invariance under the relevant symmetry group
and not as a natural or most simple choice as sometimes
advocated~\cite{Tannor97}. From a physical point of view the
master equation is expected to reflect the relevant symmetry of the
reservoir the microscopic system is interacting with. Considering for
example a single mode (harmonic oscillator) interacting with the
electromagnetic field, one has a $\mathrm{U(1)}$ symmetry and condition
(\ref{1}) is actually equivalent to the rotating wave approximation,
essentially saying that the master equation is invariant under the
transformation
\begin{displaymath}
{\hat {{\sf a}}}\rightarrow{\hat {{\sf a}}} e^{-i\theta}
\qquad
 {{\hat {{\sf a}}}^{\scriptscriptstyle\dagger}}\rightarrow {{\hat
     {{\sf a}}}^{\scriptscriptstyle\dagger}}e^{+i\theta} 
\end{displaymath}
or equivalently in terms of ${\hat {\sf x}}$ and
${\hat {\sf p}}$
\begin{equation}
  \label{2}
  {\hat {\sf x}}\rightarrow {\hat {\sf x}}\cos\theta+{\hat {\sf
      p}}\sin\theta\qquad{\hat {\sf p}}\rightarrow- {\hat {\sf
      x}}\sin\theta+{\hat {\sf p}}\cos\theta .
\end{equation}
The rotating wave approximation is therefore strictly linked to a
$\mathrm{U(1)}$ symmetry and one cannot expect or try to obtain
translational invariance in this case~\cite{reply}.
\par
If now one further makes the choice $H_0 (\hat{\sf N})=\hbar\omega
(\hat{\sf N}+\frac{1}{2})$ corresponding to a single mode (harmonic
oscillator) and asks that an operator with the canonical structure
${\hat \rho}_0=e^{-\beta H_0 (\hat{\sf N})}$ be a stationary solution,
i.e., ${\cal M}_{{\hat {{\sf a}}}{\hat
  {{\sf a}}}^{\scriptscriptstyle\dagger}}[{\hat \rho}_0]=0$, a
further connection between the coefficients of the master equation
appears~\cite{Sandulescu}
\begin{displaymath}
  2 \frac{D_{pp}l^2}{\hbar^2}=\gamma \coth\left(\frac 12
    \beta\hbar\omega\right), 
\end{displaymath}
where $\beta$ is the inverse temperature characterizing the thermal
electromagnetic field the mode is interacting with. The requirements
of complete positivity, shift-covariance and existence of the expected
canonical stationary solution then fix the quantum Fokker-Planck
equation to be of the form
\begin{eqnarray}
\label{qo}
  {\cal M}_{{\hat {{\sf a}}}{\hat
  {{\sf a}}}^{\scriptscriptstyle\dagger}}^{\rm \scriptscriptstyle
    QO}[{\hat \rho}] 
  &=&
        -
        {i\over\hbar}
        [
        {H_0 }(\hat{\sf N}),{\hat \rho}
        ]
\\
\nonumber
&\hphantom{=}&
        +\gamma
\left[\coth\left(\frac 12 \beta\hbar\omega\right)+1\right]
\left[ 
{\hat {{\sf a}}}{\hat
  \rho}{{\hat {{\sf a}}}^{\scriptscriptstyle\dagger}} -
\frac 12
        \left \{
                {\hat \rho},{{\hat {{\sf
                        a}}}^{\scriptscriptstyle\dagger}}{\hat {{\sf
                      a}}} 
        \right \}
\right]
\\
\nonumber
&\hphantom{=}&
        +\gamma
\left[\coth\left(\frac 12 \beta\hbar\omega\right)-1\right]
\left[ 
{{\hat {{\sf a}}}^{\scriptscriptstyle\dagger}}{\hat
  \rho}{\hat {{\sf a}}} -         \frac 12
        \left \{
                {\hat \rho},{\hat {{\sf a}}}{{\hat {{\sf
                        a}}}^{\scriptscriptstyle\dagger}} 
        \right \}
\right],
\end{eqnarray}
or in terms of the average of the number operator over a thermal
distribution
\begin{displaymath}
  N_\beta (\omega)=\frac{1}{e^{\beta\hbar\omega}-1}=\frac
  12\left[\coth\left(\frac 12 \beta\hbar\omega\right)-1\right] 
\end{displaymath}
setting $\eta=2\gamma$
\begin{displaymath}
  {\cal M}_{{\hat {{\sf a}}}{\hat
  {{\sf a}}}^{\scriptscriptstyle\dagger}}^{\rm \scriptscriptstyle
    QO}[{\hat \rho}] 
  =
        -
        {i\over\hbar}
        [
        {H_0 }(\hat{\sf N}),{\hat \rho}
        ]
        +\eta
\left(N_\beta (\omega)+1\right)
\left[ 
{\hat {{\sf a}}}{\hat
  \rho}{{\hat {{\sf a}}}^{\scriptscriptstyle\dagger}} -         
\frac 12
        \left \{
                {\hat \rho},{{\hat {{\sf
                        a}}}^{\scriptscriptstyle\dagger}}{\hat {{\sf
                      a}}} 
        \right \}
\right]
        +\eta N_\beta (\omega)
\left[ 
{{\hat {{\sf a}}}^{\scriptscriptstyle\dagger}}{\hat
  \rho}{\hat {{\sf a}}} -         \frac 12
        \left \{
                {\hat \rho},{\hat {{\sf a}}}{{\hat {{\sf
                        a}}}^{\scriptscriptstyle\dagger}} 
        \right \}
\right],
\end{displaymath}
i.e., the well-known quantum optical master equation for the
description of a damped harmonic oscillator (for a recent review
see~\cite{EnglertQO}), where the only free parameter is the decay rate
$\eta$ and a further freedom appears in the commutator term, where a
function of $\hat{\sf N}$ corresponding to a frequency shift may be
considered.  The quantum optical master equation is therefore
essentially fixed by formal requirements, well in accordance with its
stability with respect to microphysical derivations, which are in fact
predictive in so far as they give explicit expressions for $\eta$ and
the energy shift. As a last step, using as natural length of the
problem $l=\sqrt{\frac{\hbar}{M\omega}}$, (\ref{qo}) may be written in
terms of ${\hat {\sf x}}$ and ${\hat {\sf p}}$ as
\begin{eqnarray*}
  {\cal M}_{{\hat {{\sf x}}}{\hat {{\sf p}}}}^{\rm \scriptscriptstyle
    QO}[{\hat \rho}] 
  &=&
        -
        {i\over\hbar}
        \left[
        {
        {\hat {{\sf p}}}^2
        \over
        2M
        }+\frac 12 M\omega^2{\hat {{\sf x}}}^2,{\hat \rho}
        \right]
\\
\nonumber
&\hphantom{=}&
-
        {i\over\hbar}\frac{\gamma}{2}\left( [{\hat {{\sf x}}},\{ {\hat
            {{\sf p}}},{\hat \rho} 
        \} ] - [{\hat {{\sf p}}},\{ {\hat {{\sf x}}},{\hat \rho}
        \} ]
\right)
\\
\nonumber
&\hphantom{=}&
-
        {1\over\hbar}\frac{\gamma}{2}
\coth\left(\frac 12 \beta\hbar\omega\right)
\left(
M\omega\left[  
        {\hat {{\sf x}}},
        \left[  
        {\hat {{\sf x}}},{\hat \rho}
        \right]  
        \right] +\frac{1}{M\omega} \left[  
        {\hat {\sf p}},
        \left[  
        {\hat {\sf p}},{\hat \rho}
        \right]  
        \right] 
\right),
\end{eqnarray*}
where invariance under (\ref{2}) can be easily checked. 
\par
That $\mathrm{U(1)}$ symmetry or shift-covariance may lead under
suitable restrictions to the quantum optical master equation can also
be seen considering the recently obtained most general structure of a
proper generator of a shift-covariant completely positive quantum
dynamical semigroup given in~\cite{HolevoJFA-HolevoRMP33}, where also
the unboundedness of the operators appearing in the formal Lindblad
structure has been taken in due account, using the notion of
form-generator. The formal operator expression associated to the
form-generator is:
\begin{eqnarray}
\label{holevosc}
{\cal L}[{\hat \rho}]&=&-{i \over \hbar}
        \left[
        H (\hat{\sf N})
        ,{\hat \rho}
        \right]
+
\left[
A_{0}(\hat{\sf N}){\hat \rho}A^{\dagger}_{0}(\hat{\sf N})
       -
        \frac 12
        \left \{{\hat \rho},
       A^{\dagger}_{0}(\hat{\sf N})A_{0}(\hat{\sf N})
        \right \}
\right]
\\ \nonumber
&\hphantom{=}&
+
\sum_{m=1}^{\infty}
\left[
{\hat{\sf W}}^{\dagger}{}^{m}A_{-m}(\hat{\sf N}){\hat
  \rho}A^{\dagger}_{-m}(\hat{\sf N}){\hat{\sf W}}^{m} 
       -
        \frac 12
        \left \{{\hat \rho},
       A^{\dagger}_{-m}(\hat{\sf N}){\hat{\sf W}}^{m}{\hat{\sf
           W}}^{\dagger}{}^{m}A_{-m}(\hat{\sf N})  
        \right \}
\right]\\ \nonumber
&\hphantom{=}&
+
\sum_{m=1}^{\infty}
\left[
{\hat{\sf W}}^{m}A_{m}(\hat{\sf N}){\hat \rho}A^{\dagger}_{m}(\hat{\sf
  N}){\hat{\sf W}}^{\dagger}{}^{m} 
       -
        \frac 12
        \left \{{\hat \rho},
       A^{\dagger}_{m}(\hat{\sf N})A_{m}(\hat{\sf N}) 
        \right \}
\right]
,
\end{eqnarray}
where $H (\cdot)=H^* (\cdot)$, $A_i (\cdot)$ are functions of the
number operator, and ${\hat{\sf
    W}}=\sum_{n=0}^{\infty}|n+1\rangle\langle n|$ it the so-called
shift-operator~\cite{HolevoOLD}. Recalling the polar decompositions
${{\hat {{\sf a}}}^{\scriptscriptstyle\dagger}}={\hat{\sf
    W}}\sqrt{\hat{\sf N}+1}$ and ${\hat {{\sf a}}}={\hat{\sf
    W}}^{\dagger}\sqrt{\hat{\sf N}}$, for the following simple choice
of functions:
\begin{eqnarray*}
  H (n)&=& H_0 (n)=\hbar\omega \left(n+\frac 12\right)
\\   
 A_m (n)&=&0 \quad m=0, |m|>1
\\
A_1 (n)&=&\sqrt{\gamma (\beta)}\sqrt{n+1}  
\\ 
A_{-1} (n)&=&e^{\frac{\beta}{2}}\sqrt{\gamma (\beta)}\sqrt{n} ,
\end{eqnarray*}
where
\begin{displaymath}
\gamma(\beta)
=
\gamma
\left[\coth\left(\frac 12
    \beta\hbar\omega\right)-1\right]  ,
\end{displaymath}
one recovers from (\ref{holevosc}) the quantum optical
master equation.
\par
It is interesting to observe that complete positivity,
shift-covariance and the requirement of a canonical stationary
solution also allow as a proper generator of a quantum dynamical
semigroup the following expression
\begin{eqnarray}
\label{pippo}
  {\cal L}[{\hat \rho}]  &=&
        -
        {i\over\hbar}
        [
        {H_0 }(\hat{\sf N}),{\hat \rho}
        ]
-\gamma_0        \left[  
        {\hat {{\sf N}}},
        \left[  
        {\hat {{\sf N}}},{\hat \rho}
        \right]  
        \right]  
\\
\nonumber
&\hphantom{=}&
        +\sum_{m=1}^{\infty}\gamma_m
\left\{
\left[\coth\left(\frac 12 \beta\hbar\omega\right)+1\right]^m
\left[ 
{\hat {{\sf a}}}^m{\hat
  \rho}{{\hat {{\sf a}}}^{\scriptscriptstyle\dagger}}{}^m -         
\frac 12
        \left \{
                {\hat \rho},{{\hat {{\sf
                        a}}}^{\scriptscriptstyle\dagger}}{}^m{\hat
                  {{\sf a}}}^m 
        \right \}
\right]\right.
\\
\nonumber
&\hphantom{=}&
\left.
\hphantom{+\sum_{m=1}^{\infty}\gamma_m\left[\right.}
+
\left[\coth\left(\frac 12 \beta\hbar\omega\right)-1\right]^m
\left[ 
{{\hat {{\sf a}}}^{\scriptscriptstyle\dagger}}{}^m{\hat
  \rho}{\hat {{\sf a}}}^m -         \frac 12
        \left \{
                {\hat \rho},{\hat {{\sf a}}}^m{{\hat {{\sf
                        a}}}^{\scriptscriptstyle\dagger}}{}^m 
        \right \}
\right]
\right\},
\end{eqnarray}
corresponding to the functions:
\begin{eqnarray*}
A_0 (n)&=&\sqrt{\gamma_0}n   
\\ A_m (n)&=&\sqrt{\gamma_m (\beta)}
\sqrt{\frac{(n+m)!}{n!}}
\\ A_{-m}
(n)&=&e^{m\frac{\beta}{2}}\sqrt{\gamma_m (\beta)} 
\sqrt{\frac{n!}{(n-m)!}}  
,  
\end{eqnarray*}
where
\begin{displaymath}
\gamma_m(\beta) 
=
\gamma_m
\left[\coth\left(\frac 12
    \beta\hbar\omega\right)-1\right]^m  .
\end{displaymath}
Eq.~(\ref{pippo}) provides a generalization of the quantum optical
master equation in which both phase-diffusion, related to the
coefficient $\gamma_0$, and $m$-photon processes with decay rates
$\eta_m=2^m\gamma_m$, which should quickly approach zero in order to
allow for non-explosion of the associated Markov process, can be
considered. Thus far we have dealt with the case of shift-covariance,
corresponding to $\mathrm{U(1)}$ symmetry of the system with many
degrees of freedom acting as reservoir and determining the
non-Hamiltonian dynamics of the microsystem.
\subsection{Translation-covariance}
\label{IB}
We now consider the case in which the relevant symmetry is invariance
under translations, corresponding to a homogeneous reservoir. Given
the unitary representation of the translation group
\begin{displaymath}
  \hat{U} (b) =e^{-{i\over\hbar}b{\hat {\sf p}}},
\end{displaymath}
where ${\hat {\sf p}}$ is the momentum operator of the microsystem and
$b\in\mathrm{R}$, translation-covariance according
to~\cite{HolevoRAN-HolevoRMP32} amounts to the requirement
\begin{equation}
  \label{200}
    {\cal L}_{{\hat {{\sf x}}}{\hat {{\sf
            p}}}}[e^{-{i\over\hbar}b{\hat {\sf
          p}}}\cdot e^{+{i\over\hbar}b{\hat {\sf
          p}}}]=e^{-{i\over\hbar}b{\hat 
        {\sf p}}}{\cal L}_{{\hat {{\sf x}}}{\hat {{\sf
            p}}}}[\cdot]e^{+{i\over\hbar}b{\hat {\sf p}}}.  
\end{equation}
Invariance under the group $\mathrm{R}$ of translations thus implies 
for the structure of the quantum Fokker-Planck equation that the
Hamiltonian has to be a function of the generator of the
transformation ${\hat {\sf p}}$ and the following simple requirement
in the coefficients appearing in (\ref{xp})
\begin{displaymath}
  \mu=\gamma.
\end{displaymath}
Considering for example a free particle interacting with
a homogeneous reservoir, one has $\mathrm{R}$ symmetry
corresponding to invariance under translations, which is
reflected by the fact that according to (\ref{200}) the
master equation is invariant under the transformation
\begin{displaymath}
  {\hat {\sf x}}\rightarrow {\hat {\sf x}}+b\qquad {\hat {\sf
      p}}\rightarrow {\hat {\sf p}}
\end{displaymath}
or equivalently in terms of ${\hat {{\sf a}}}$ and ${{\hat {{\sf
        a}}}^{\scriptscriptstyle\dagger}}$ 
\begin{displaymath}
  {\hat {{\sf a}}}\rightarrow {\hat {{\sf
        a}}}+\frac{1}{\sqrt{2}}\frac{b}{l}
\qquad {{\hat {{\sf
        a}}}^{\scriptscriptstyle\dagger}}\rightarrow {{\hat {{\sf
        a}}}^{\scriptscriptstyle\dagger}}+\frac{1}{\sqrt{2}}\frac{b}{l}.
\end{displaymath}
Further making the obvious choice $H_0 ({\hat {{\sf p}}})={{\hat {{\sf
        p}}}^2 \over 2M}$, corresponding to a free particle of mass
$M$ and asking that an operator with the canonical structure ${\hat
  \rho}_0=e^{-\beta H_0 ({\hat {{\sf p}}})}$ be a stationary solution,
i.e., ${\cal L}_{{\hat {{\sf x}}}{\hat {{\sf p}}}}[{\hat \rho}_0]=0$,
one has the condition
\begin{displaymath}
  D_{pp}=\gamma \frac{2M}{\beta},
\end{displaymath}
with $\beta$ the inverse temperature of the homogeneous reservoir. The
requirement of complete positivity, translation-covariance and
existence of the expected stationary solution thus constrain the
quantum Fokker-Planck equation to be of the form
\begin{eqnarray}
  \label{qbm}
  {\cal L}_{{\hat {{\sf x}}}{\hat {{\sf p}}}}^{\rm \scriptscriptstyle
    QBM}[{\hat \rho}] 
  &=&
        -
        {i\over\hbar}
        [
        {H_0 ({\hat {{\sf p}}})},{\hat \rho}
        ]
        -
        {i\over\hbar}\gamma[ {\hat {{\sf x}}},\{{\hat
          {{\sf p}}},{\hat \rho} \} ] 
        -\gamma
        {
        2M
        \over
         \beta\hbar^2
        }
        \left[  
        {\hat {{\sf x}}},
        \left[  
        {\hat {{\sf x}}},{\hat \rho}
        \right]  
        \right]  
\\
\nonumber
&\hphantom{=}&
        -
        {
        D_{xx}
        \over
         \hbar^2
        }
        \left[  
        {\hat {\sf p}},
        \left[  
        {\hat {\sf p}},{\hat \rho}
        \right]  
        \right]  
        +2
        {
        D_{px}  
        \over
         \hbar^2
        }
        \left[  
        {\hat {{\sf x}}},
        \left[  
        {\hat {{\sf p}}},{\hat \rho}
        \right]  
        \right]          
 ,
\end{eqnarray}
i.e., the typical structure one finds in the extensive physical
literature aiming at the description of quantum Brownian motion
(see~\cite{SandulescuIJMPE,Isar99} for a review). With respect to the
case of the quantum optical master equation (\ref{qo}) the remaining
freedom in the structure is much bigger: apart from the friction
coefficient $\gamma$ and a real function of ${\hat {{\sf p}}}$
correcting the Hamiltonian, the coefficients $D_{xx}$ and $D_{px}$ are
undetermined except for the relation
\begin{displaymath}
   \gamma
        {
        2M
        \over
         \beta
        } D_{xx}-D_{px}^2\geq \frac{\gamma^2\hbar^2}{4}
\end{displaymath}
stemming from (\ref{vincoli}). This is reflected by the fact that a much wider literature has
been devoted to the subject, looking for microphysical derivations of
the quantum Brownian motion master equation, in order to obtain
expressions for the undetermined parameters.
\par
We now compare the result (\ref{qbm}) with the general structure of a
proper generator of a translation-covariant quantum dynamical
semigroup, given in~\cite{HolevoRAN-HolevoRMP32}, where also the case
of an unbounded generator has been considered, introducing the notion
of form-generator and specifying a suitable domain for the mapping.
Restricting to the continuous component of the generator,
corresponding to a quantum Fokker-Planck equation describing friction
and diffusion, one has for the formal operator expression associated
to the form-generator the following result
\begin{equation}
  \label{holevotc}
  {\cal L}
[{\hat \rho}]
  =
        -
        {i\over\hbar}
        [
        {\beta{\hat {{\sf x}}}+ H ({\hat {{\sf p}}})},{\hat \rho}
        ]
        +\alpha {\hat {{\sf V}}}{\hat \rho}{\hat {{\sf V}}}^{\dagger}
        -{\hat {{\sf K}}}{\hat \rho}-
        {\hat \rho}{\hat {{\sf K}}}^{\dagger}
        \qquad \beta\in \mathrm{R}, \ \alpha \geq 0
\end{equation}
\begin{displaymath}
{\hat {{\sf V}}}={\hat {{\sf x}}}+L({\hat {{\sf p}}}) \qquad {\hat
  {{\sf K}}}=\frac{\alpha}{2}
\left[{\hat {{\sf x}}}^2+2{\hat {{\sf x}}}L({\hat {{\sf
        p}}})+L^{\dagger}({\hat {{\sf p}}})L({\hat {{\sf p}}})
\right], 
\end{displaymath}
$H (\cdot)=H^* (\cdot)$, $L (\cdot)$ being functions of the momentum
operator ${\hat {{\sf p}}}$, and $\beta\neq 0$ implying e.g., a
constant gravitational or electric field. According to
(\ref{holevotc}) one has a single operator of the form ${\hat {{\sf
      V}}}={\hat {{\sf x}}}+L({\hat {{\sf p}}})$ (or one for each
Cartesian coordinate considering higher dimensions) instead of two as
considered in (\ref{l}). Expressing (\ref{holevotc}) in terms of
nested commutators and anticommutators as in (\ref{xp}), restricting
to the case in which $L (\cdot)$ is a linear function, according to
the fact that we are considering friction effects at most linear in
the velocity, the inequalities in (\ref{vincoli}), corresponding to
the fact that the determinant of the matrix given in (\ref{matrix}) be
zero or positive, now become more restrictive:
\begin{displaymath}
  D_{xx}\geq 0 \qquad D_{pp}\geq 0 \qquad D_{xx}D_{pp}-D_{px}^2=
  \frac{\gamma^2\hbar^2}{4}, 
\end{displaymath}
corresponding to $\det { \mathbf{D}}=0$.
Coming back to (\ref{qbm}) this implies the further restriction
\begin{displaymath}
  D_{xx}=\gamma \frac{\beta\hbar^2}{8M}+\frac{\beta}{2\gamma M}D_{px}^2,
\end{displaymath}
so that apart from the overall multiplying coefficient $\gamma$ only
another coefficient $D_{px}$ is left free, and one has
\begin{eqnarray}
  \label{300}
  {\cal L}_{{\hat {{\sf x}}}{\hat {{\sf p}}}}^{\rm \scriptscriptstyle
    QBM}[{\hat \rho}] 
  &=&
        -
        {i\over\hbar}
        [
        {H_0 ({\hat {{\sf p}}})},{\hat \rho}
        ]
        -
        {i\over\hbar}\gamma[ {\hat {{\sf x}}},\{{\hat
          {{\sf p}}} ,{\hat \rho} \}] 
        -\gamma
        {
        2M
        \over
         \beta\hbar^2
        }
        \left[  
        {\hat {{\sf x}}},
        \left[  
        {\hat {{\sf x}}},{\hat \rho}
        \right]  
        \right]  
        -\gamma
        {
        \beta
        \over
        8M
        }
        \left[  
        {\hat {\sf p}},
        \left[  
        {\hat {\sf p}},{\hat \rho}
        \right]  
        \right]  
\\
\nonumber
&\hphantom{=}&
        -
        {
        \beta
        \over
        2\gamma M
        }
        {
        D_{px}^2  
        \over
         \hbar^2
        }
        \left[  
        {\hat {{\sf p}}},
        \left[  
        {\hat {{\sf p}}},{\hat \rho}
        \right]  
        \right]  
        +2
        {
        D_{px}
        \over
         \hbar^2
        }
        \left[  
        {\hat {\sf p}},
        \left[  
        {\hat {\sf x}},{\hat \rho}
        \right]  
        \right]  .
\end{eqnarray}
Therefore a predictive microphysical model of quantum Brownian motion
essentially has to indicate an explicit expression for the
coefficients $\gamma$ and $D_{px}$.
\par
It is interesting to express (\ref{300}) in terms of the creation and
annihilation operators given in (\ref{relazione}). Setting ${
  \beta\hbar^2 \over 4M }=\lambda^2_{\rm \scriptscriptstyle th}$, the
square of the thermal wavelength of the microsystem undergoing
Brownian motion, one has:
\begin{eqnarray}
\label{400}
  {\cal L}_{{\hat {{\sf a}}}{\hat
  {{\sf a}}}^{\scriptscriptstyle\dagger}}^{\rm \scriptscriptstyle
    QBM}[{\hat \rho}] 
  &=&
        -
        {i\over\hbar}
        [
        {H_0 ({\hat {{\sf a}}},{\hat
  {{\sf a}}}^{\scriptscriptstyle\dagger})},{\hat \rho}
        ]
        -
        \frac{\gamma}{2}[{\hat {{\sf a}}}^2-{\hat
  {{\sf a}}}^{\scriptscriptstyle\dagger}{}^2 ,{\hat \rho}]  
\\
\nonumber
&\hphantom{=}&
        +
\left[ \frac{\gamma}{2}
\left(\frac{\lambda^2_{\rm \scriptscriptstyle
      th}}{l^2}+\frac{l^2}{\lambda^2_{\rm \scriptscriptstyle
      th}}+2\right)+\frac{1}{\gamma}
\frac{2}{\hbar^2}\frac{\lambda^2_{\rm \scriptscriptstyle th}}{l^2}
D_{px}^2  
\right]
\left[ 
{\hat {{\sf a}}}{\hat
  \rho}{{\hat {{\sf a}}}^{\scriptscriptstyle\dagger}} -
\frac 12
        \left \{
                {\hat \rho},{{\hat {{\sf
                        a}}}^{\scriptscriptstyle\dagger}}{\hat {{\sf
                      a}}} 
        \right \}
\right]
\\
\nonumber
&\hphantom{=}&
        +
\left[ \frac{\gamma}{2}
\left(\frac{\lambda^2_{\rm \scriptscriptstyle
      th}}{l^2}+\frac{l^2}{\lambda^2_{\rm \scriptscriptstyle
      th}}-2\right)+\frac{1}{\gamma}
\frac{2}{\hbar^2}\frac{\lambda^2_{\rm \scriptscriptstyle th}}{l^2}
D_{px}^2  
\right]
\left[ 
{{\hat {{\sf a}}}^{\scriptscriptstyle\dagger}}{\hat
  \rho}{\hat {{\sf a}}} -         \frac 12
        \left \{
                {\hat \rho},{\hat {{\sf a}}}{{\hat {{\sf
                        a}}}^{\scriptscriptstyle\dagger}} 
        \right \}
\right]
\\
\nonumber
&\hphantom{=}&
        -
\left[ \frac{\gamma}{2}
\left(\frac{\lambda^2_{\rm \scriptscriptstyle
      th}}{l^2}-\frac{l^2}{\lambda^2_{\rm \scriptscriptstyle
      th}}\right)+ 
2\frac{D_{px}}{\hbar}\left(\frac{1}{\gamma}\frac{\lambda^2_{\rm
      \scriptscriptstyle th}}{l^2}\frac{D_{px}}{\hbar} -i\right) 
\right]
\left[ 
{\hat {{\sf a}}}{\hat
  \rho}{\hat {{\sf a}}} -         \frac 12
        \left \{
                {\hat \rho},{\hat {{\sf a}}}^2
        \right \}
\right]
+\mathrm{h.c.}
\end{eqnarray}
Eq.~(\ref{400}) strongly simplifies if one takes for the length $l$,
used in order to introduce the operators ${\hat {{\sf a}}}$ and
${{\hat {{\sf a}}}^{\scriptscriptstyle\dagger}}$ in terms of the
operator position and momentum of the particle, the value
$l=\lambda_{\rm \scriptscriptstyle th}=\sqrt{{ \beta\hbar^2 \over 4M
    }}$, naturally suggested by the underlying physics. With $l$ the
thermal de Broglie wavelength (\ref{400}) reduces to
\begin{eqnarray}
\label{finale}
  {\cal L}_{{\hat {{\sf a}}}{\hat
  {{\sf a}}}^{\scriptscriptstyle\dagger}}^{\rm \scriptscriptstyle
    QBM}[{\hat \rho}] 
  &=&
        -
        {i\over\hbar}
        [
        {H_0 ({\hat {{\sf a}}},{\hat
  {{\sf a}}}^{\scriptscriptstyle\dagger})},{\hat \rho}
        ]
        -
        \frac{\gamma}{2}[{\hat {{\sf a}}}^2-{\hat
  {{\sf a}}}^{\scriptscriptstyle\dagger}{}^2 ,{\hat \rho}]  
        +
2\gamma
\left[ 
{\hat {{\sf a}}}{\hat
  \rho}{{\hat {{\sf a}}}^{\scriptscriptstyle\dagger}} -         
\frac 12
        \left \{
                {\hat \rho},{{\hat {{\sf
                        a}}}^{\scriptscriptstyle\dagger}}{\hat {{\sf
                      a}}} 
        \right \}
\right]
\\
\nonumber
&\hphantom{=}&
+ \frac{2}{\gamma}
\frac{D_{px}^2}{\hbar^2}
\left[ 
{\hat {{\sf a}}}{\hat
  \rho}{{\hat {{\sf a}}}^{\scriptscriptstyle\dagger}} -         
\frac 12
        \left \{
                {\hat \rho},{{\hat {{\sf
                        a}}}^{\scriptscriptstyle\dagger}}{\hat {{\sf
                      a}}} 
        \right \}
+
{{\hat {{\sf a}}}^{\scriptscriptstyle\dagger}}{\hat
  \rho}{\hat {{\sf a}}} -         \frac 12
        \left \{
                {\hat \rho},{\hat {{\sf a}}}{{\hat {{\sf
                        a}}}^{\scriptscriptstyle\dagger}} 
        \right \}
\right]
\\
\nonumber
&\hphantom{=}&
        -
2\frac{D_{px}}{\hbar}\left(\frac{1}{\gamma}
\frac{D_{px}}{\hbar} -i\right) 
\left[ 
{\hat {{\sf a}}}{\hat
  \rho}{\hat {{\sf a}}} -         \frac 12
        \left \{
                {\hat \rho},{\hat {{\sf a}}}^2
        \right \}
\right]
+\mathrm{h.c.}
,
\end{eqnarray}
where the last three contributions can only vanish if the real
coefficient $D_{px}$ is equal to zero, corresponding to $L (\cdot)=-L^*
(\cdot)$ in (\ref{holevotc}). Eq.~(\ref{300}) or equivalently
(\ref{finale}) express the general structure of a quantum
Fokker-Planck equation which is invariant under translation, warrants
the existence of the expected canonical expression as a stationary
solution, thus recovering equipartition, and is furthermore a proper
generator of a completely positive quantum dynamical semigroup.
\subsection{Covariance and uniqueness of the stationary solution}
\label{IC}
In recent work~\cite{Tannor97,Yan} aiming at comparing and clarifying
different approaches to quantum dissipation, which relies
on~\cite{LindbladQBM} but neglects the more recent and thorough
results of~\cite{HolevoRAN-HolevoRMP32,HolevoJFA-HolevoRMP33}, the
statement can be found that \textit{no Markovian theory can satisfy
  all three criteria of positivity, translational invariance, and
  asymptotic approach to the canonical equilibrium state $e^{-\beta
    H_0}$, except in special cases}. This statement is always correct
in view of the last observation, and these simple exceptional cases,
which are usually not spelled out, can be read in (\ref{holevosc}):
the microsystem has to be a free particle apart from an effective
correction to the Hamiltonian, given by a real function of ${\hat
  {{\sf p}}}$ describing for example an effective mass, and a
potential term depending linearly on position (such as, e.g., a
constant gravitational or electric field). These cases are often
neglected, having in mind that translational invariance, mainly seen
as an abstract property rather than the expression of homogeneity of
the reservoir, should be always asked for.  In this spirit
translational invariance, i.e., $\mathrm{R}$ symmetry, is asked for
also for the damped harmonic oscillator.
\par
The three physical requirements one can reasonably ask together for
Markovian systems in the weak-coupling limit are complete positivity,
existence of the stationary solution predicted by equipartition and
invariance under the relevant symmetry, not necessarily translational
invariance.  Having translational invariance apart from the potential
term is not physically significant since the potential term actually
breaks this invariance, and would furthermore lead to high
non-uniqueness of the stationary state, as argued below. The dynamics
of the microsystem is driven by both the potential term (which could
also arise as a mean field effect) and the contributions describing
decoherence and dissipation, so that a physically relevant symmetry
should pertain to the system as a whole.  Let us in fact consider a
mapping ${\cal F}$ covariant with respect to a given symmetry group
$G$ according to (\ref{100}), which admits a stationary solution
${\hat \rho}_0$, i.e., ${\cal F}[{\hat \rho}_0]=0$. If the operator
${\hat \rho}_0$ is not invariant under the unitary representation
${\cal U}_g$ of $G$, so that
\begin{displaymath}
  {\hat \rho}_g={\cal U}_g[{\hat \rho}_0]
\end{displaymath}
is linearly independent from ${\hat \rho}_0$ at least for some $g$ in
$G$, due to the $G$-covariance of ${\cal F}$, ${\hat \rho}_g$ still
is a stationary solution
\begin{displaymath}
  {\cal F}[{\hat \rho}_g]={\cal F}[{\cal U}_g[{\hat \rho}_0]]={\cal
    U}_g[{\cal F}[{\hat \rho}_0]]=0,  
\end{displaymath}
so that one cannot have the expected uniqueness of the solution. In
the case of the one-dimensional Lie groups considered in Sec.~\ref{II}
for example a stationary solution can be unique only if it commutes
with the generator of the group. Note that this simple argument is
independent on whether the mapping ${\cal F}$ ensures complete
positivity of the time evolution or not, so that the clash between the
requirement of translation invariance and the existence of the correct
stationary state corresponding to equipartition is not due to the
requirement of complete positivity of the time evolution mapping.
\section{QUANTUM FOKKER-PLANCK EQUATION FOR THE MOTION OF A BROWNIAN
  PARTICLE}
\label{III}
As we have shown in Sec.~\ref{IA}, in the case of an underlying
$\mathrm{U(1)}$ symmetry formal requirements are enough to essentially
fix structure and coefficients of the quantum Fokker-Planck equation,
apart from an energy shift and an overall coefficient. The same is not
true in the case of $\mathrm{R}$ symmetry describing translational
invariance. In this paragraph we therefore quickly recall a recently
proposed quantum Fokker-Planck equation for the description of the
motion of a heavy Brownian particle interacting through collisions
with a homogeneous fluid made up of much lighter particles
(see~\cite{art3,art4,art5} for details). This result has been obtained
within a kinetic approach, where the dynamics is driven by single
events described as collisions in which one generally has momentum and
energy transfer, alternative to the Zwanzig Caldeira Leggett
approach~\cite{Zwanzig-Caldeira} (recently criticized
in~\cite{AlickiCL}), where one describes the reservoir as a collection
of harmonic oscillators coupled to the microsystem through its
position, usually performing calculations in terms of path-integral
techniques.
\par
In the aforementioned approach one obtains in the first instance a
kinetic equation for the statistical operator analogous to the
classical linear Boltzmann equation given by the simple expression
\begin{eqnarray}
  \label{general}
         {\cal L} [{\hat \rho}] 
        =
        &-&
        {i \over \hbar}
        \left[
        H_0 ({\hat {{\sf p}}}) 
        ,
        {\hat \rho}
        \right]
      \nonumber
      \\
        &+&
        {2\pi \over\hbar}
        (2\pi\hbar)^3
        n
        \int_{{\bf R}^3} d^3\!
        {{\bf q}}
        \,  
        {
        | \tilde{t} (q) |^2
        }
      \Biggl[
        e^{{i\over\hbar}{{\bf q}}\cdot{\hat {{\sf x}}}}
        \sqrt{
        S({{\bf q}},{\hat {{\sf p}}})
        }
        {\hat \rho}
        \sqrt{
        S({{\bf q}},{\hat {{\sf p}}})
        }
        e^{-{i\over\hbar}{{\bf q}}\cdot{\hat {{\sf x}}}}
        -
        \frac 12
        \left \{
        S({{\bf q}},{\hat {{\sf p}}}),
        {\hat \rho}
        \right \}
        \Biggr],
\end{eqnarray}
where $ \tilde{t} (q)$ is the Fourier transform of the T-matrix
describing the microphysical collisions and the function $S({\bf{q}},
{\bf{p}})$ appearing operator-valued in (\ref{general}) is a positive
two-point correlation function known in the physical community as
dynamic structure factor~\cite{Lovesey}, usually expressed as a
function of momentum and energy transfer, ${\bf{q}}$ and $E$,
according to
\begin{displaymath}
S({\bf{q}},{\bf{p}})\equiv  S({\bf{q}},E) \qquad  E ({\bf{q}},{\bf{p}})=
{
({\bf{p}}+{\bf{q}})^2
\over
   2M
}
-
{
{\bf{p}}^2
\over
   2M
},
\end{displaymath}
with $M$ the mass of the Brownian particle. The dynamic structure
factor $S({\bf{q}},E)$ is the Fourier transform of the two-point time
dependent density autocorrelation function of the fluid
\begin{equation}
  \label{dsf}
  S({\bf{q}},E) = 
        {  
        1  
        \over  
         2\pi\hbar
        }  
        \int_{{\bf R}} dt 
        {\int_{{\bf R}^3} d^3 \! {\bf{x}} \,}        
        e^{
        {
        i
        \over
         \hbar
        }
        [E ({\bf{q}},{\bf{p}}) t -
        {\bf{q}}\cdot{\bf{x}}]
        } 
      \frac{1}{N}
        {\int_{{\bf R}^3} d^3 \! {\bf{y}} \,}
        \left \langle  
         N({\bf{y}})  
         N({\bf{y}}+{\bf{x}},t)  
        \right \rangle        ,
\end{equation}
and is always positive since it is proportional to the energy
dependent scattering cross-section of a microscopic probe off a
macroscopic sample~\cite{vanHove}, giving the spectrum of its
spontaneous fluctuations.  Equation (\ref{general}) also has the three
properties of complete positivity, translational invariance and
canonical stationary solution, and it actually gives a physical
example of the general structure of generator of a
translation-covariant quantum dynamical
semigroup~\cite{HolevoRAN-HolevoRMP32}, going beyond the diffusive
case considered in (\ref{holevosc}). In order to recover from the
general integral kinetic equation (\ref{general}) a quantum
Fokker-Planck equation for the description of the Brownian motion of a
heavy particle one has to consider the limit of small momentum
transfer and small energy transfer (corresponding to the Brownian
limit in which the mass of the test particle is much heavier than the
particles of the fluid).  Considering a gas of free particles obeying
Maxwell-Boltzmann statistics, due to
\begin{equation}
  \label{7}
        S_{\rm \scriptscriptstyle MB}({\bbox{q}},{\bbox{p}})
        =
        {
        1
        \over
         (2\pi\hbar)^3
        }
        {
        2\pi m^2
        \over
        n\beta q
        }
        z
        \exp\left[
        -{
        \beta
        \over
             8m
        }
        {
        (2mE({\bbox{q}},{\bbox{p}}) + q^2)^2
        \over
                  q^2
        }
        \right]
\end{equation}
one has~\cite{art5}
\begin{equation}
        \label{qbmnostra}
          {\cal L} [{\hat \rho}]
        =
        -
       {i\over\hbar}
        [
        H_0 ({\hat {{\sf p}}}) 
        ,{\hat \rho}
        ]
        -
        {i\over\hbar}
        \gamma
        \sum_{i=1}^3
        \left[  
        {\hat {{\sf x}}}_i ,
        \left \{  
        {\hat {\sf p}}_i,{\hat \rho}
        \right \}  
        \right]      
        -
        {
        D_{pp}  
        \over
         \hbar^2
        }
        \sum_{i=1}^3
        \left[  
        {\hat {{\sf x}}}_i,
        \left[  
        {\hat {{\sf x}}}_i,{\hat \rho}
        \right]  
        \right]  
        -
        {
        D_{xx}
        \over
         \hbar^2
        }
        \sum_{i=1}^3
        \left[  
        {\hat {\sf p}}_i,
        \left[  
        {\hat {\sf p}}_i,{\hat \rho}
        \right]  
        \right].  
\end{equation}
In this kinetic case the free parameters in (\ref{300}) are now
determined as:
\begin{displaymath}
  D_{px}=0\qquad \gamma=\frac 13 z {\pi^2 m^2 \over\beta\hbar}        
\int_{{\bf R}^3} d^3\!
        {{\bf q}}
        \,  
        {
        | \tilde{t} (q) |^2
        }
        q
        e^{-
        {
        \beta
        \over
             8m
        }
        {{{q}}^2}
        }
\end{displaymath}
with $z=e^{\beta\mu}$ the fugacity of the gas~\cite{Huang}, while
$D_{xx}$ and $D_{pp}$ are expressed in terms of $\gamma$ as can be
read in (\ref{300})
\begin{displaymath}
  D_{xx}={
        \beta\hbar^2
        \over
         8M
        } \gamma
\qquad
D_{pp}={
        2M
        \over
         \beta
        }\gamma.
\end{displaymath}
Also the expression of (\ref{qbmnostra}) in terms of the operators
${\hat {{\sf a}}}_i$ and ${{\hat {{\sf
        a}}}_i^{\scriptscriptstyle\dagger}}$ according to
(\ref{finale}) takes in this case a particularly simple form
\begin{equation}
\label{last}
  {\cal L}[{\hat \rho}]
  =
        -
        {i\over\hbar}
        [
        {H_0 ({\hat {{\sf a}}}_i,{\hat
  {{\sf a}}}_i^{\scriptscriptstyle\dagger})},{\hat \rho}
        ]
        -
        \frac{\gamma}{2}\sum_{i=1}^3 [{\hat {{\sf a}}}_i^2-{\hat
  {{\sf a}}}^{\scriptscriptstyle\dagger}_i{}^2 ,{\hat \rho}]  
        +
2\gamma\sum_{i=1}^3
\left[ 
{\hat {{\sf a}}}_i{\hat
  \rho}{{\hat {{\sf a}}}^{\scriptscriptstyle\dagger}}_i -
\frac 12
        \left \{
                {\hat \rho},{{\hat {{\sf
                        a}}}^{\scriptscriptstyle\dagger}}_i{\hat {{\sf
                      a}}}_i 
        \right \}
\right]
,
\end{equation}
so that one has a single ${\hat {{\sf
      a}}}_i=\sqrt{\frac{2M}{\beta\hbar^2}} ({\hat {{\sf
      x}}}_i+i\frac{\beta\hbar^2}{4M} {\hat {{\sf p}}}_i)$ operator
for each Cartesian coordinate.
\section{CONCLUSIONS AND OUTLOOK}
\label{IV}
The main scope of this paper was to show how relevant symmetries can
be in the determination of structures of quantum Fokker-Planck
equation. While in the literature only translational invariance is
considered and asked for also in the case of the damped harmonic
oscillator, leading to no go statements regarding the possibility of
having quantum Fokker-Planck equations with all the physically
relevant features (complete positivity, covariance and equipartition),
we here considered both one-dimensional Lie groups $\mathrm{U(1)}$ and
$\mathrm{R}$, showing that they are connected to two distinct classes
of quantum Fokker-Planck equations: the quantum optical
master equation associated to shift-covariance and the quantum
Brownian motion master equation associated to translation-covariance.
That these two classes of models actually correspond to different
physics can also be seen in connection with recent studies on their
properties with respect to decoherence~\cite{HuPRA-Kofman}.  In
particular, independently of complete positivity of the time
evolution, it has been shown that covariance properties put severe
restrictions on the structure of the stationary solution, provided
uniqueness is asked for.
\par
In the case of shift-covariance a generalization of the quantum
optical master equation has been proposed, in which also $m$-photon
processes can be considered.  Moreover recent mathematical results by
Holevo have been considered, concerning the structure of generators of
completely positive quantum dynamical semigroups and leading to
further restrictions in the quantum Brownian motion case. For the
Brownian motion of a test particle in a fluid, where formal
requirements are not enough to essentially fix the structure of the
quantum Fokker-Planck equation describing the phenomenon, a recent
microphysical approach has been briefly recalled, based on scattering
theory, where the quantum Fokker-Planck equation is obtained as the
small momentum and energy transfer limit of a quantum generalization
of the classical linear Boltzmann equation.  Covariance properties
with respect to some physically relevant group, typically reflecting a
symmetry under certain transformations of the given reservoir, can
therefore be a most useful requirement in the determination of
structures of quantum Fokker-Planck equations or more generally of
linear kinetic equations.
\par
An interesting extension of this work would entail the study of
generators of the dynamics of systems in which one has an important
correlation between internal and translational degrees of freedom,
both coupled through different interactions to some reservoir, a
problem recently considered in~\cite{AlickiBB}, where the
translational degrees of freedom are treated in a classical way,
assuming decoherence is strong enough. Such models could be of
interest for the implementation of quantum computing, where indeed
some experimental scheme actually relies on a coupling between
internal and center of mass degrees of freedom~\cite{ZeilingerQC}.
\section*{ACKNOWLEDGMENTS}
The author would like to thank Prof. L. Lanz for many useful
suggestions during the whole work and Prof. A.  Barchielli for useful
discussions. He also thanks Dr.~F.~Belgiorno for careful reading of
the manuscript. This work was partially supported by MIUR under
Cofinanziamento and Progetto Giovani.
 
\end{document}